\documentclass[onecolumn]{article}
\usepackage{subfigure}
\usepackage{amsmath}
\usepackage{amssymb}
\usepackage[utf8]{inputenc}
\usepackage[lmargin=3.5cm,rmargin=3.5cm,tmargin=3.5cm,bmargin=4.5cm]{geometry}
\usepackage{placeins}
\usepackage{caption}
\usepackage{subfigure}
\usepackage{dblfloatfix}
\usepackage{braket}
\usepackage{multicol}
\usepackage{bm}
\usepackage{graphicx}
\usepackage{titletoc}

\title{Analysis of the Breakdown of Exponential Decays of Resonances}
\author{Adam Wyrzykowski}
\date{\today}
\begin{document}

\maketitle

\section*{Abstract}
In the paper, a simple model of alpha decay with Dirac delta potential is studied.
The model leads to breakdown of the exponential decay and to power law behavior
at asymptotic times. Time dependence of the survival probability of the particle in
the potential well is analyzed numerically with two methods: integration of Green's function representation and numerical solution of the time-dependent Schr\"{o}dinger equation. In particular, finite depth potential wells and behavior between the exponential and power law regimes, which are situations that could not be described in detail analytically, are studied. The numerical results confirm power law with exponent $n=3$ after the turnover into the non-exponential decay regime. Moreover,
the constructive and destructive interference is observed in the intermediate stage
of the process. The simple alpha decay model is compared to the results of Rothe-
Hintschich-Monkman experiment which was the first experimental proof of violation
of the exponential law.

\section{Introduction}

Exponential law in alpha decay was first explained by Gamow in 1928 \cite{gamow}. In his reasoning, eigenfunctions with complex eigenenegies are present, which initially caused some concerns about validity of this approach. Nevertheless, this method leads to correct predictions about the exponential time dependence and the Geiger-Nutall law. Although the exponential decay law provides a very good description for quasi-stationary states, it is only an approximate solution. It was observed by Khalfin in 1958 \cite{khalfin}, that the exponential behavior cannot hold for all times $t\in(0,\infty)$. In particular, for asymptotic times $t\to\infty$, the survival probability $P(t)$ of a state decreases slower than any exponential function \cite{khalfin, urbanowski}:
\begin{equation}
P(t)\leq A\text{e}^{-bt^q},
\end{equation}
if the energy distribution density is bounded from below or above ($A>0$, $b>0$ and $0<q<1$).\\
Since then, several models of systems that exhibit violation of the
exponential law were developed (see e.g. \cite{cavalcanti, krainov}). Among the physical phenomena described by these models one can find: spontaneous decay in two-level systems, alpha-decay, single photon ionization of atoms, etc. Assuming a certain model of a phenomenon, it can be shown that the survival probability obeys power law for sufficiently large times \cite{khalfin, cavalcanti, krainov, rothe}:
\begin{equation}
P(t)\sim\begin{cases}
\text{e}^{-t/\tau};& \text{ for } t\lesssim t_{\text{breakdown}}\\
t^{-n};& \text{ for } t\gtrsim t_{\text{breakdown}}.
\end{cases}
\end{equation}
Its exponent $n$ depends on the phenomenon and the model, typically taking values between $1$ and $4$.\\
Deviation from exponential behavior can be understood by considering the time evolution of a state \cite{khalfin, urbanowski, giacosa}:
\begin{equation}
\ket{\psi(t)}=\int_{-\infty}^{\infty}dE\ e^{-iEt/\hslash} c(E) \ket{E},
\end{equation}
where $\ket{E}$ are eigenstates of the Hamiltonian. The amplitude that the initial state has not decayed yet at time $t$ is:
\begin{equation}
a(t)=\braket{\psi(0)|\psi(t)}=\int_{-\infty}^{\infty}dE\ e^{-iEt/\hslash}\omega(E).
\end{equation}
Here, $\omega(E)=c^*(E)c(E)$ is the energy distribution density. If the energy density is modeled by the Breit-Wigner distribution:
\begin{equation}
\omega(E)=\frac{1}{\pi}\frac{\Gamma}{(E-E_0)^2+\Gamma^2},
\label{5}
\end{equation}
the amplitude equals
\begin{equation}
a(t)=\exp\left(-\frac{-iE_0t}{\hslash}-\frac{\Gamma|t|}{\hslash}\right).
\end{equation}
Thus, in this case, the time dependence of the survival probability is given by an exponential function. However, the energy density must be bounded from below in order for the ground state to exist. The distribution (\ref{5}) is therefore approximate, some corrections in $\omega(E)$ must be included and new terms will also appear in the formulas for the amplitude and the probability. These terms decrease slower than the exponential function and become dominant for large times \cite{khalfin}.\\

Although the power law time evolution is expected at large times, its experimental observation is not trivial. If the exponential stage of a process lasts 20 or more lifetimes, the survival probability may be too low to be measured. This would require accuracy of the order $e^{-20}\approx10^{-9}$ or better, but no high energy resonance is measured with such precision. Therefore, discovery of decay processes which cease to follow the exponential behavior sufficiently early is a major theoretical and experimental challenge. The first experimental proof of the turnover into the non-exponential decay regime, based on the measurements of the luminescence decays of dissolved organic materials, was found only quite recently, namely in 2006 \cite{rothe}. Another proof of non-exponential behavior at large times, yet indirect, was found in the scattering process $\alpha^+\alpha\to^8\text{Be}(2^+)\to\alpha^+\alpha$ \cite{nowakowski}. It is indirect in a sense that the energy density was measured, the difference from the Breit-Wigner distribution was observed, but the survival probability was calculated based on these data.\\

The paper is organized as follows. Section II provides an analysis of the simple, one-dimensional model of alpha decay from \cite{cavalcanti}. Section III includes numerical results for this model in situations, which are not well described analytically. In particular, the intermediate stage of the process between the exponential and power law regimes, and the case, when a potential strength $\lambda$ does not satisfy the condition $\lambda\gg1$, were studied. These calculations are compared to the results of Rothe-Hintschich-Monkman experiment \cite{rothe}. The final remarks and discussion are presented in section IV.

\begin{figure*}[t]
\begin{center}
\includegraphics[width=12cm]{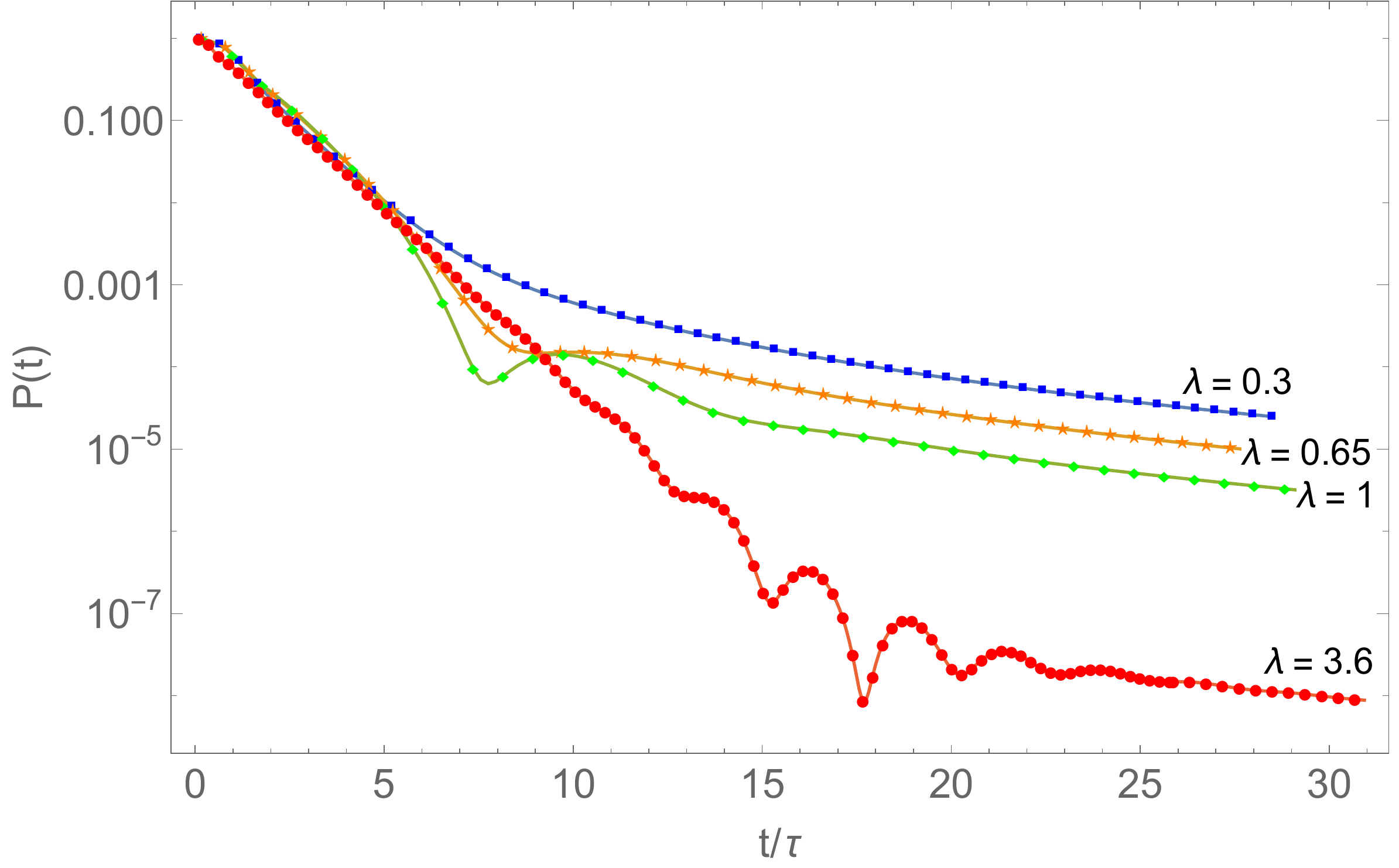}
\end{center}
\caption{\small{Time dependence of the survival probability for $\lambda=0.3,\ 0.65,\ 1,\ 3.6$, logarithmic plot. The horizontal axis is chosen such that the initial slope of all curves is $-1$, i.e. the time is measured as a ratio $t/\tau$, where $\tau$ is the lifetime of a corresponding state.}}
\end{figure*}

\section{Model}

The alpha-decay model studied in this paper assumes a potential:
\begin{equation}
V(x)=\begin{cases}
\frac{\lambda\hslash^2}{2ma}\delta(x-a)& \text{ for } x>0,\\
+\infty& \text{ for } x\leq 0.
\label{7}
\end{cases}
\end{equation}
Analysis of this potential can be found in \cite{cavalcanti}. Following Gamow's reasoning \cite{gamow}, one could use an outgoing wave ansatz in the region outside the potential well, $\varphi_2(x)=Be^{ikx}$ for $x>a$. Such an assumption leads to a discrete spectrum of complex eigenenergies and the eigenfunctions which decay exponentially in time. The most troublesome corollary of this approach is the fact that the eigenfuctions diverge when $x\to\infty$, so are not normalizable. The way out of this problem is to treat the process as a scattering one, i.e. an incoming wave must be included in the solution for the outer region, $\varphi_2(x)=e^{-ikx}+Be^{ikx}$ for $x>a$. It can be easily seen that, taking into account the boundary conditions, the most general form of the solution inside the potential well is $\varphi_1(x)=A \sin kx$ ($0<x<a$). From the matching conditions at $x=a$, one finds that the spectrum of eigenenergies is continuous and real. The coefficients $A=A(k)$ and $B=B(k)$ are:
\begin{equation}
A(k)=-\frac{2ika}{ka+\lambda e^{ika}\sin ka},
\label{8}
\end{equation}
\begin{equation}
B(k)=-\frac{ka+\lambda e^{-ika}\sin ka}{ka +\lambda e^{ika}\sin ka}.
\end{equation}
The wavefunction inside the well can be expressed as \cite{cavalcanti}:
\begin{equation}
\psi(x,t)=\frac{1}{2\pi}\int_0^\infty e^{-\frac{i\hslash}{2m}k^2t}\phi(k)|A(k)|^2\sin kx\ dk,
\label{10}
\end{equation}
where
\begin{equation}
\phi(k)\equiv\int_0^a \psi(x',0)\sin kx'\ dx'
\end{equation}
depends on the initial wavefunction. When $\lambda\gg1$, the poles of analytic continuation of $|A(k)|^2$, which is:
\begin{equation}
W(k)=-A(-k)A(k)=\frac{4k^2a^2}{(ka)^2+\lambda ka\sin(2ka)+\lambda^2\sin^2(ka)},
\label{12}
\end{equation}
are very close to the real axis. Thus, the leading contribution to the integral (\ref{10}) comes from points on the real axis that lie in the vicinity of poles. To proceed, one would like to expose the role of poles and get rid of the oscillatory factor $e^{-(i\hslash/2m)k^2t}$ in eq. (\ref{10}). This can be achieved by shifting the integration contour. After rotation in the clockwise sense by $45^\circ$, one obtains:
\begin{equation}
\psi(x,t)=e^{-i\pi/4}\int_0^\infty e^{-\frac{\hslash k^2}{2m}t}f(e^{-i\pi/4}k,x)\ dk +\sum_{n=1}^\infty C(k_n,x)e^{-\frac{i\hslash}{2m}k_n^2t},
\label{13}
\end{equation}
where
\begin{equation}
f(k,x)\equiv \frac{1}{2\pi}\phi(k)W(k)\sin kx
\end{equation}
and $-C(k_n,x)/2\pi i$ are contributions from the poles of $f(k,x)$ at $k=k_n$. In the above formula, the first term has power behavior at large times, while the second one corresponds to exponential decays. The survival probability, i.e. the probability that the particle remains in the potential well, is defined as $P(t)=\int_0^a|\psi(x,t)|^2 dx$. When the background integral in eq. (\ref{13}) can be neglected, the survival probability is roughly equal to:
\begin{equation}
P_{\text{poles}}(t)\approx \sum_{n=1}^\infty c_ne^{-\Gamma_n t/\hslash},
\label{15}
\end{equation}
where $c_n=\int_0^a|C(k_n,x)|^2 dx$ and $\Gamma_n=-\text{Im}(\hslash^2 k_n^2/m)$. However, for sufficiently large times, the contribution of the background integral to the survival probability follows the power law:
\begin{equation}
P_{\text{background}}(t)\sim\frac{m^3a^6}{\lambda^4t^3}
\end{equation}
and eventually dominates over the pole contribution. The time when it happens, i.e. the time of the breakdown of the exponential decay, can be estimated to be $t_\text{breakdown}\sim 10 (\hslash/\Gamma_1)\ln \lambda$ \cite{cavalcanti}.\\

The problem of a particle in the potential (\ref{7}) is related to the case of an infinite potential well. When $\lambda\gg 1$, the two problems are almost the same and, as can be seen from eq. (\ref{12}), the poles correspond to the momenta which satisfy $\sin(k_n a)\approx 0$. This is the well-known condition in the case of the infinite well. Thus, there is a correspondence between the n-th excited state:
\begin{equation}
\psi^{(n)}(x)=N\sin\left(\frac{n\pi x}{a}\right)
\end{equation}
and the n-th pole (or the n-th contributions in the sums in eqs. (\ref{13}) and (\ref{15})). This relation should also hold approximately for finite $\lambda$'s. In particular, if the initial wavefunction is chosen to be $\psi(x',0)=\psi^{(n)}(x')$, $c_n$ is expected to be the largest coefficient in the sum (\ref{15}). The values of the weights $c_n$ when $\psi^{(n)}(x)$ is used as the initial condition, with $n=1,\ 2,\ 3,\ 4$, are presented in Table 1.
\begin{table}[h!]
\begin{center}
\begin{tabular}{|r||c|c|c|c|c|}
\hline
$n$&$c_1$&$c_2$&$c_3$&$c_4$&$c_5$\\ \hline
1&1.012&0.022&0.005&0.002&0.001\\ \hline
2&0.016&1.059&0.060&0.013&0.006\\ \hline
3&0.005&0.036&1.148&0.106&0.023\\ \hline
4&0.003&0.013&0.057&1.270&0.157\\
\hline
\end{tabular}
\caption{\small{Dependence of the value of weights $c_n$ on the number $n$ of the excited state $\psi^{(n)}$ used as the initial condition ($\lambda=8$).}}
\end{center}
\end{table}\\

All above statements will hold in 3 dimensions, for $\ell=0$. This is because the radial equation for $rR(r)$ ($R(r)$ -- radial part of the wavefunction), when $\ell=0$, is the same with the time-independent Schr\"{o}dinger equation in 1 dimension, $rR(r)\to0$ at $r=0$ \cite{messiah}, the matching conditions are clearly the same and, when calculating the probability, the extra $r^{-2}$ factor cancels out with $r^2$ from the Jacobian.

\section{Consequences of the model}
\subsection{Numerical results}
To find the time dependence of the survival probability, the wavefunction is calculated from eq. (\ref{13}) at 100 equally spaced points from $x=0$ to $x=a$ and then the survival probability, $P(t)=\int_0^a |\psi(x,t)|^2\ dx$ is obtained by the trapezoid method (to simplify the calculations). The initial wavefunction $\psi(x',0)=\sqrt{2/a}\sin(\pi x'/a)$ is used. This method, based on integration of the Green's function representation, was compared with a direct numerical solution of the time-dependent Schr\"{o}dinger equation for times of the order of 1-2 lifetimes. In the latter approach, a Gaussian potential barrier of width $\Delta$, centered around $x=a$ was used instead of a Dirac delta potential to have a smooth function. Then, the results were extrapolated to $\Delta\to 0$. The two methods provide compatible results, e.g. the difference between their predictions for $|\psi(0.6a,\ 0.4\tau_0)|^2$ is about $0.1\ \%$ (when $\lambda=8$). Here, $\tau_0\equiv m(\lambda a)^2/(2\pi^3\hslash)$ is a characteristic time unit of the process.  For larger times, in particular when the breakdown of the exponential decay occurs, only the Green's function approach was used.
\begin{figure}
\centering
\subfigure[$\lambda=1$, logarithmic plot]{
\setbox1=\hbox{\includegraphics[height=8cm]{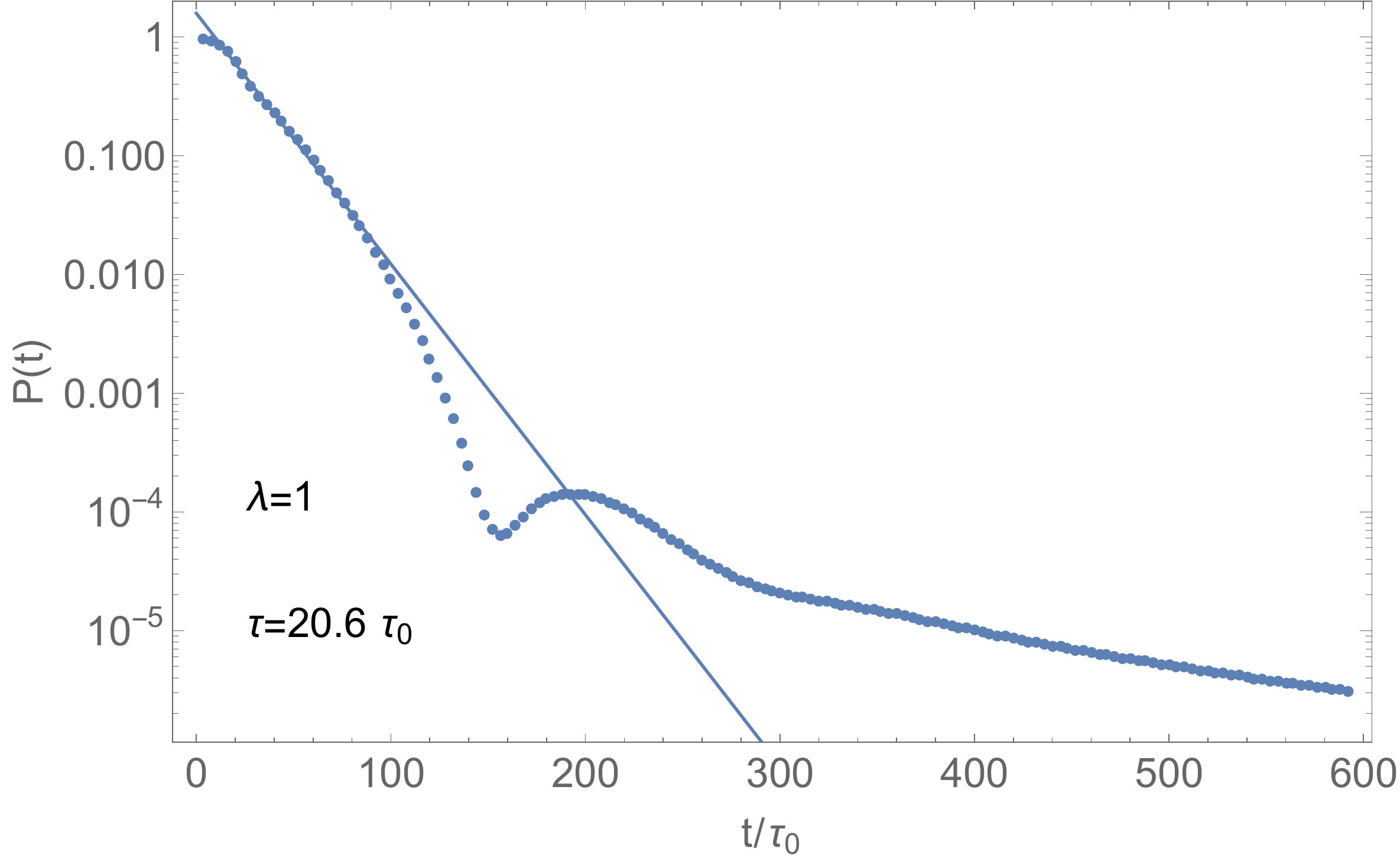}}
 \includegraphics[height=7cm]{2a.pdf}\llap{\raisebox{3.4cm}{\includegraphics[height=3.5cm]{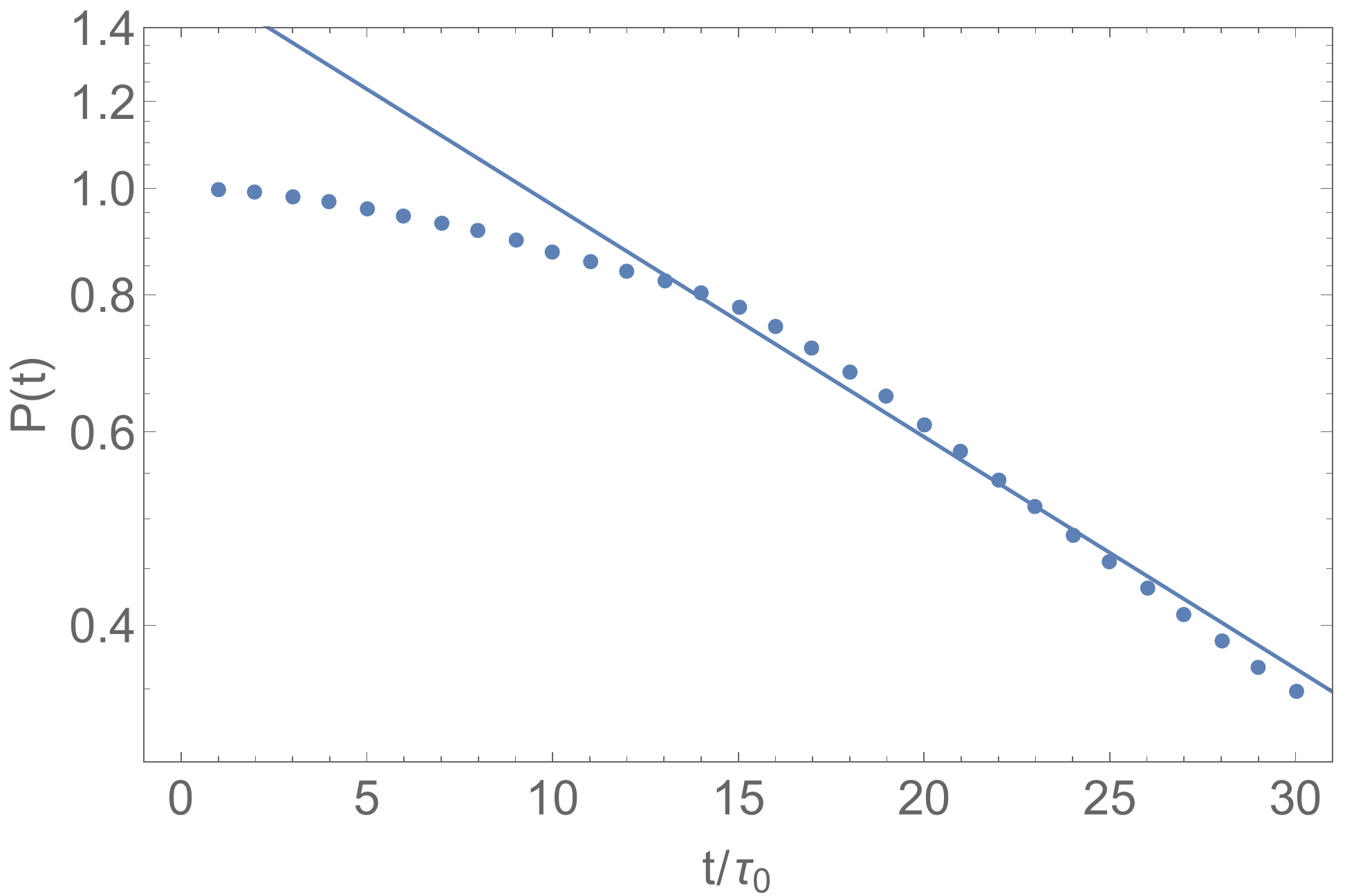}\quad }}
\label{2a}}
\subfigure[$\lambda=1$, log-log plot]{
\includegraphics[height=7cm]{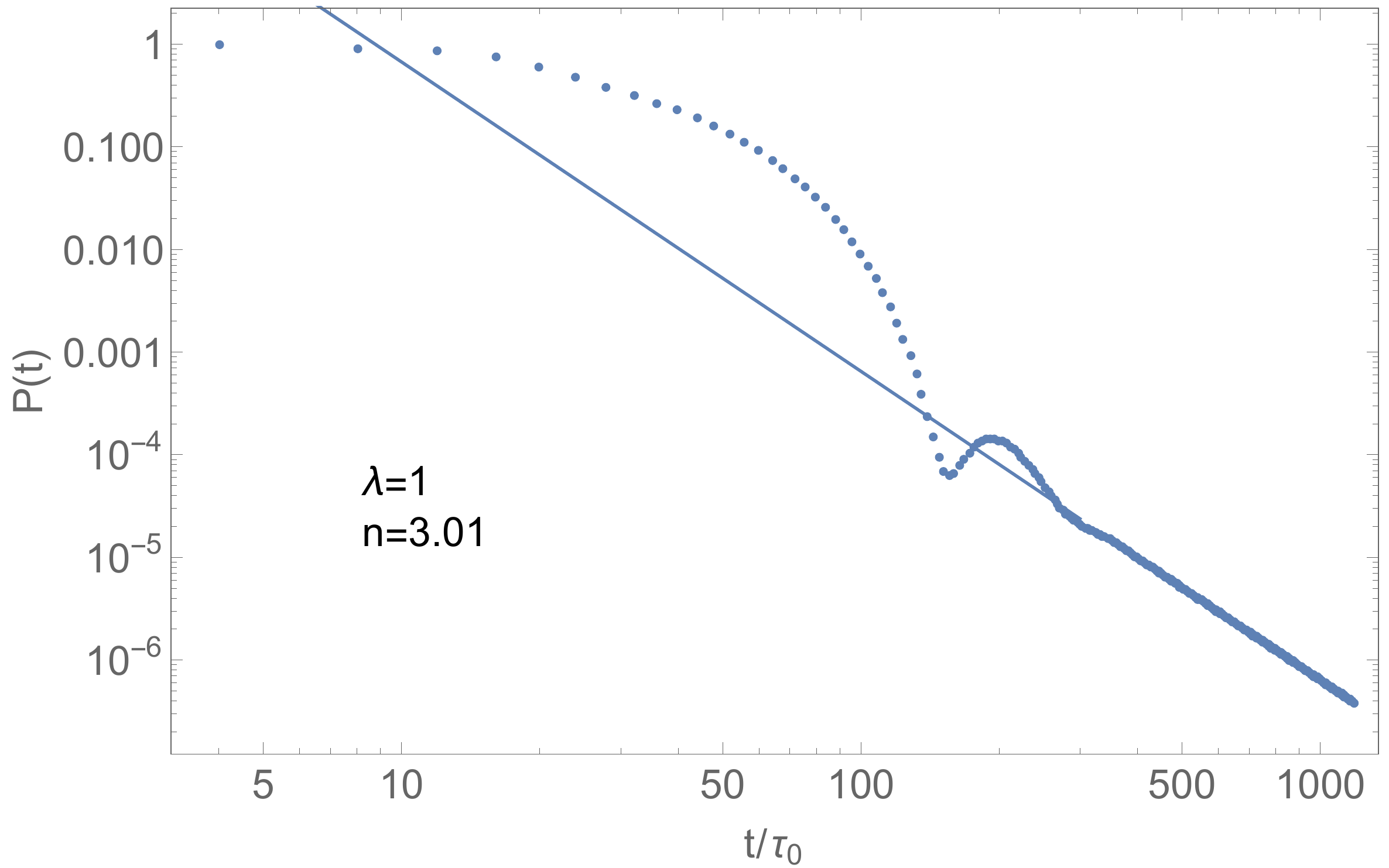}
\label{2b}}
\caption{\small{Time dependence of the survival probability. One can observe exponential behavior in the early stage of the process, deviation from exponential function very close to $t=0$ and power time dependence at large times. Exponential and power functions are fitted in appropriate regimes (solid lines).}}
\end{figure}\\

The time dependence of the survival probability is presented in Figure 1. The results of the fit of the exponential function to the early time behavior and the power law function for large times are shown in Figure 2 for $\lambda=1$. The time axis in Figure 1 is chosen such that the initial slope of all curves is $-1$, i.e. the time is measured as a ratio $t/\tau$, where $\tau$ is the lifetime of a corresponding state, which is found from the fit.\\
Both the exponential and power law regimes can be easily observed, so the theoretical predictions from \cite{cavalcanti} are confirmed in these numerical calculations. The exponent $n$ of the power law $P(t)=Bt^{-n}$ agrees with the theoretical value $n=3$ within uncertainty of the fit, as shown in Table 2.
\begin{table}
\begin{center}
\begin{tabular}{|c|c|}
\hline
$\lambda$&$n$\\
\hline
$0.3$&$3.010\pm0.017$\\
\hline
$0.65$&$2.996\pm0.020$\\
\hline
$1$&$2.992\pm0.012$\\
\hline
$3.6$&$3.000\pm0.040$\\
\hline
\end{tabular}
\end{center}
\caption{\small{Power law exponent fitted in the power law regime of the survival probability time dependence for various $\lambda$.}}
\end{table}\\
Also the fact that the behavior for very small times is not exponential, but initially $\dot{P}(0)=0$, is observed (see the inset in Fig. 2a). The phenomenon that the decay of a quasi-stationary state is not exponential in this regime is known from the theory \cite{cavalcanti, misra} and was observed in experiment \cite{wilkinson}. It is related to the \emph{quantum Zeno paradox}.\\

A new effect, which was noticed, is the presence of a sigle ($\lambda=1$) or many ($\lambda=3.6$) oscillations of the survival probability. They occur in the intermediate stage of the process, between the exponential and power law regimes. In this region, the survial probability is not a monotonic function of time. Such behavior is caused by interference between the background integral part of eq. (\ref{13}) and the pole term when they are of comparable order. If one splits the wavefunction as in eq. (\ref{13}):
\begin{equation}
\psi(x,t)=\psi_{\text{bg}}(x,t)+\psi_{\text{poles}}(x,t),
\end{equation}
not only the total survival probability can be calculated, but also quantities:
\begin{equation}
P_{\text{bg}}(t)=\int_0^a |\psi_{\text{bg}}(x,t)|^2\ dx
\end{equation}
\begin{equation}
P_{\text{poles}}(t)=\int_0^a|\psi_{\text{poles}}(x,t)|^2\ dx
\end{equation}
\begin{equation}
P_{\text{interf}}(t)=\int_0^a\left(\psi^*_{\text{bg}}(x,t)\psi_{\text{poles}}(x,t)+\text{h.c.}\right)dx.
\end{equation}
They represent the separate contributions of the background, poles and interference term to the total survival probability. The interference term has an unexpected feature -- it changes it sign abruptly. This explains better the origin of the oscillations. For early times, the pole contribution dominates, and for asymptotic times contribution from the background integral is leading. Only for intermediate times $P_{\text{interf}}(t)$ is comparable to the rest of the survival probability $P_{\text{poles}}(t)+P_{\text{bg}}(t)$. If it changes its sign in this intermediate stage, the derivative of $P(t)$ will rapidly decrease (when $P_{\text{interf}}(t)$ changes the sign from plus to minus) or increase (the opposite case). These abrupt changes of $P_{\text{interf}}(t)$ cause the oscillations. Briefly, the reason for the oscillations is constructive and destructive interference between the pole and background contributions. Another, probably more physical, way of understanding these oscillations is interference of the outgoing and incoming waves included in our solution.\\

An important question, especially from the point of view of experimental observation of the breakdown of exponential decays of resonances, is what are the parameters for which turnover to the power law regime occurs at the earliest time. A good and general parameter for this purpose is \emph{Q-value} of a resonance:
\begin{equation}
Q=\frac{\epsilon_n}{\Gamma_n}=-\frac{\text{Re }E_n}{2\text{ Im }E_n},
\end{equation}
where $\epsilon_n=\text{Re }E_n$ is the energy of a resonance and $\Gamma_n=-2\text{Im }E_n$ -- its width, i.e. $E_n=\epsilon_n-i\Gamma_n/2$. Time of the breakdown of the exponential behavior (in units of the lifetime) is proportional to the logarithm of $Q$ \cite{rothe}, thus the lower the Q-value, the earlier it happens. On the other hand, if the Q-value is too small, the resonance is not well formed. Thus, to have a well formed resonance and to observe breakdown of the exponential law before there is a very low probability that the state has not decayed yet, a compromise must be made for the optimal choice of $Q$. Another way to check whether a resonance is still well formed is to calculate the lifetime in two ways: from the fit (as in Fig. 2a) or by finding the first pole of $A(k)$, eq. (\ref{8}) and calculating the corresponding lifetime:
\begin{equation}
\tau_1=\frac{\hslash}{\Gamma_1}=-\frac{m}{\text{Im}(\hslash k_1^2)}.
\end{equation}
If the resonance is well formed and there is a single dominant pole, the two methods should provide similar results. Analysis of quality of resonances for small $\lambda$ is shown in Table 3. From these data, one can infer that the optimal $\lambda$ is roughly about $3.6$ (depending on the criterion used). For this parameter, deviations from the exponential behavior start when the signal is $P(t)/P(0)\sim 10^{-5}$ of the initial value and the system enters the power law regime for $P(t)/P(0)\sim 10^{-8}$ (see Figure 1).
\begin{table}
\begin{center}
\begin{tabular}{|r||c|c|c|c|}
\hline
$\lambda$&$Q$&$\tau/\tau_0$(fit)&$\tau_1/\tau_0$(poles)&[\%]\\ \hline
0.3&0.208&204&119&53\\ \hline
0.65&0.454&47.0&33.9&32\\ \hline
1&0.667&20.8&17.6&17\\ \hline
3.6&2.48&3.55&3.48&2.0\\ \hline
\end{tabular}
\end{center}
\caption{\small{Q-values, lifetimes from the fit, lifetimes found from the first poles and the relative difference between these two results for various $\lambda$.}}
\end{table}

\subsection{Comparison with experiment}

The main problem with observation of nonexponential regime in experiment is the fact that, in general, it occurs after many lifetimes, so the signal is too weak to be detected. Since the time of breakdown of exponential decay is lower for lower Q-values, a process with a broad energy spectrum is needed rather than a narrow, well-formed resonance. This fact made searches for non-exponential decays in nuclear and particle physics inefficient. Another physical system had to be used and the first successfull experimental observation of violation of the exponential decay law was made by Rothe, Hintschich and Monkman \cite{rothe}, in 2005. It was found in atomic data, namely in measurements of the luminescent decays of dissolved organic materials. Both the impact of the solvent environment and the fact, that large, organic molecules were used, leads to broadening of the energy spectrum. This effect is sufficient to observe the power law behavior all the way up to about 20 lifetimes (e.g. 17 for Rhodamine 6G and 11 for polyfluorene).\\

Their results can be compared to the predictions of the simple model studied here. The experimental data for Rhodamine 6G from \cite{rothe} are compared with the curves from numerical calculations for the alpha decay model in Figure 3. The value of $\lambda$, that fits the experimental results best, is:
\begin{equation}
\lambda\approx 3.6
\end{equation}
For $\lambda=3.6$, the lifetime found by fitting the exponential function to the theoretical curve is (see Table 3) $\tau_{th}\approx 3.55\ \tau_0$, where $\tau_0=m(\lambda a)^2/(2\pi^3\hslash)$. The experimental value found in \cite{rothe} reads $\tau_{exp}=3.9\ ns$. Equating $\tau_{th}=\tau_{exp}$, one obtains after simple transformations and substitution of the numerical values:
\begin{equation}
ma^2=\frac{2\pi^3\hslash}{\lambda^2}\frac{\tau_{exp}}{\tau_{th}/\tau_0}\approx 1.2\times 10^5\ m_pa_0^2,
\label{25}
\end{equation}
where $m_p$ is the mass of proton and $a_0$ is the Bohr radius. Large value of this constant, in these units, originates from the fact that the organic molecules used in the experiment are large and dissolved in the solvent, so both the mass and length scales are much larger than $m_p$ and $a_0$. There is actually a very simple way to obtain this number: the molecular mass of Rhodamine 6G ($C_{28}H_{31}N_2O_3Cl$) is $A\approx479$, the sum of atomic numbers of its constituents is $Z=254$ and $AZ=479\cdot254\approx 1.2\times 10^5$. Such a simple relation does not hold for other substances from the experiment, though. Still, even in those other cases $AZ$ provides a reasonable estimate of the order of magnitude.\\

As can be seen in Figure 3b, although the time dependence in the non-exponential region is well described by a power function $P(t)=Bt^{-n}$, its exponent $n$ is a bit larger for the experimental data than for the theoretical curves. The exponents of the power law found for various materials in Rothe-Hintschich-Monkman experiment vary from 2 to 4 \cite{rothe}, which is not far from $n=3$ predicted in the alpha decay model (especially for Coumarin 450, $n\approx 2.9$), but it also indicates deviations from this very simple model. The physics of dissolved organic materials is certainly more complicated than the simple potential description, but apparently (\ref{7}) provides a satisfactory first approximation.
\begin{figure}
\centering
\subfigure[logarithmic scale]{
\includegraphics[width=10cm]{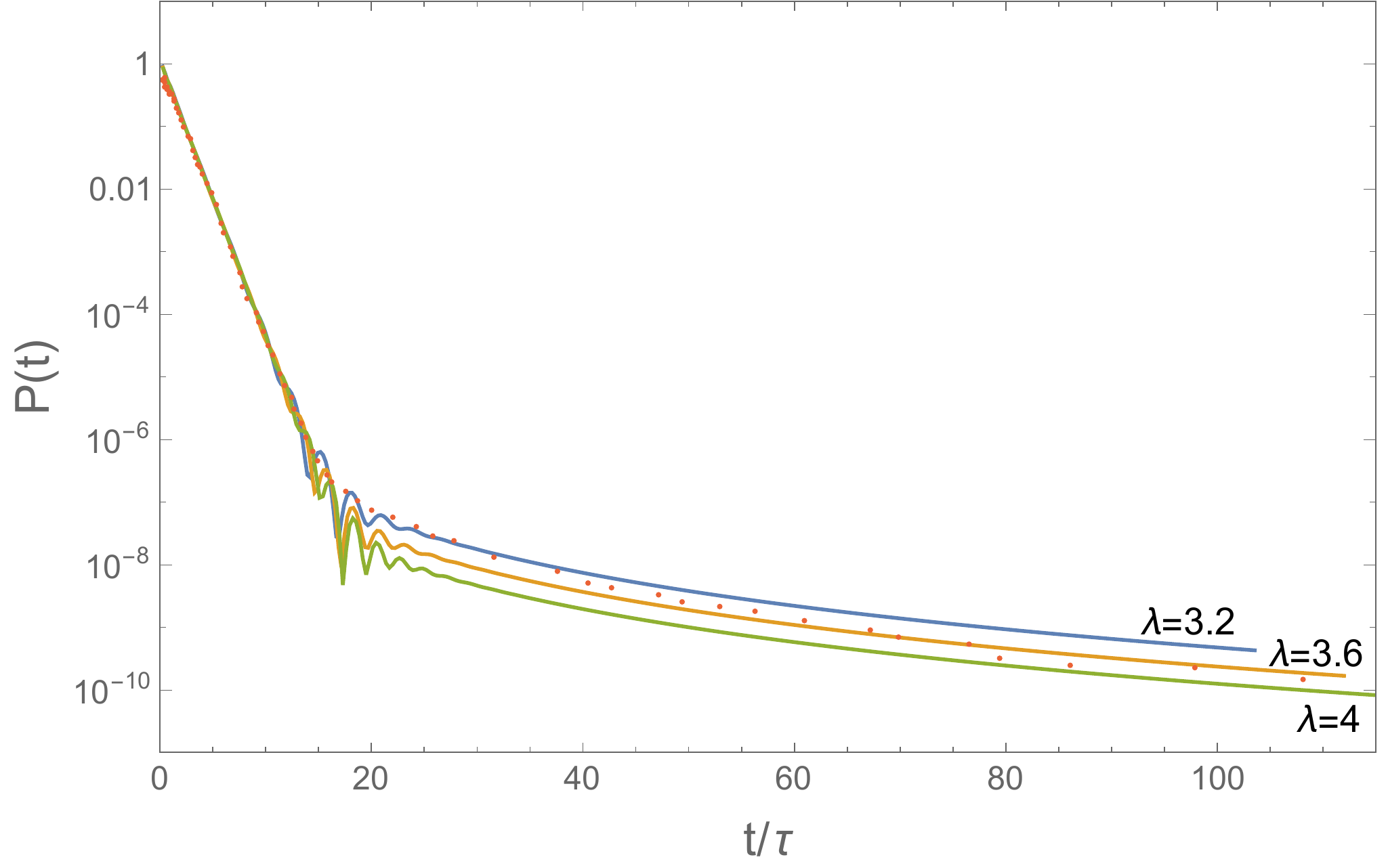}
\label{3b}}
\subfigure[log-log scale]{
\includegraphics[width=10cm]{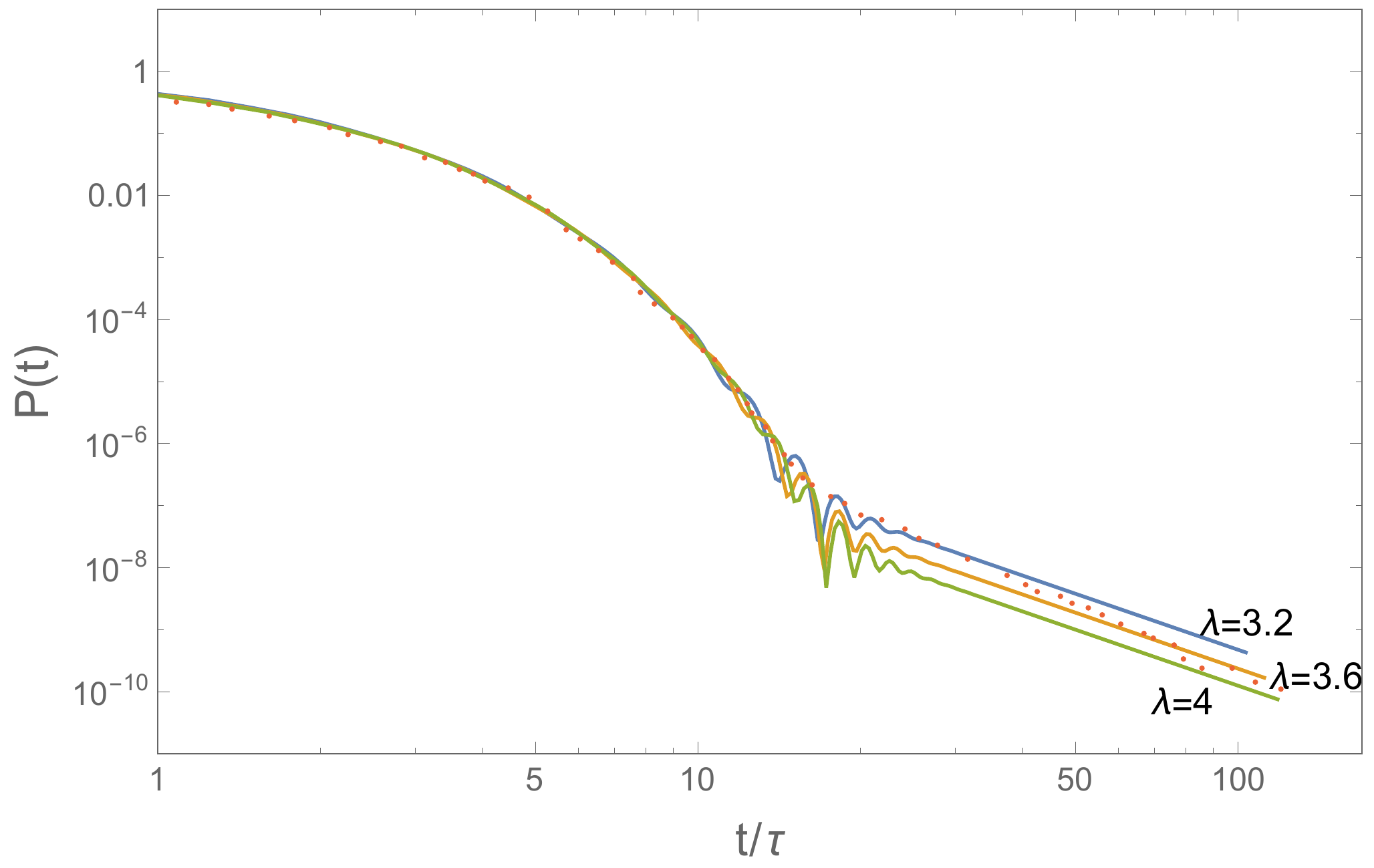}
\label{3b}}
\caption{\small{Comparison of the theoretical curves with the experimental data. The curves correspond to the time dependence of the survival probability calculated numerically for the alpha-decay model with a potential strength $\lambda=3.2,\ 3.6,\ 4$. The points are experimental results for Rhodamine 6G \cite{rothe}.}}
\end{figure}
\section{Conclusions}
In the present paper, the simple model of breakdown of exponential decay law was studied and compared for the first time with experiment. Surprisingly, the model reproduces rather well the main features of the data. In this approximation, breakdown of the initial exponential time dependence occurs and at large times, the process follows power behavior. These predictions, which originate from \cite{cavalcanti}, were tested numerically to check analytic estimates and examine the low $\lambda$ regime. The exponential and power law regimes were directly observed. The power law exponent was found to be $n=3$ within uncertainty of the fit, as predicted. The breakdown of the exponential law was seen to occur at earlier times for low $\lambda$ (low Q-value). On the other hand, high Q-values mean better formation of resonances, so from experimental point of view, a compromise must be made. Such a balanced value is about $\lambda\approx 3.6$ ($Q\approx 2.5$). Moreover, a new phenomenon was observed -- oscillation of the survival probability in the intermediate stage of the process. It originates from the constructive and destructive interference. However, the effect is likely to be model-dependent, in particular it is not observed in \cite{rothe}.\\
The results of the model studied here were compared to the experiment by Rothe, Hintschich and Monkman, which is the first experimental observation of turnover from the exponential time dependence to the power one. The simple model with a Dirac delta barrier successfully describes the experiment in both the exponential and power law regimes. To our knowledge, such quantitative comparison between theory and experiment was never attempted before. An intriguing scaling in terms of the molecular mass and the atomic number was also found.  Although behavior of dissolved organic materials is believed to be governed by much more complicated theory and some deviations from the simple model are observed, the one-dimensional model (\ref{7}) provides a good, effective description.
\subsection*{Acknowledgements}
I would like to thank Jacek Wosiek for suggesting the subject and constant encouragement, and Francesco Giacosa for instructive discussions. This work is supported by the NCN grant: UMO-2016/21/B/ST2/01492.

\end{document}